\documentclass[12pt]{article}

\usepackage{amssymb}
\usepackage{amsfonts}
\usepackage{pstricks}

\newtheorem{theorem}{Theorem}

\newtheorem{definition}[theorem]{Definition}

\begin{document}

\title{Lattice model of protein conformations}

\author{S.Albeverio\footnote{University of Bonn, Germany}, S.V.Kozyrev\footnote{Steklov Mathematical Institute, Russian Academy of Sciences}}

\maketitle

\begin{abstract}
We introduce a lattice model of protein conformations which is able to reproduce second structures of proteins (alpha--helices and beta--sheets). This model is based on the following two main ideas. First, we model backbone parts of amino acid residues in a peptide chain by edges in the cubic lattice which are not parallel to the coordinate axes. Second, we describe possible contacts of amino acid residues using a discrete model of the Ramachandran plot.

This model allows to describe hydrogen bonds between the residues in the backbone of the peptide chain. In particular the lattice secondary structures have the correct structure of hydrogen bonds. We also take into account the side chains of amino acid residues and their interaction.

The expression for the energy of conformation of a lattice protein which contains contributions from hydrogen bonds in the backbone of the peptide chain and from interaction of the side chains is proposed. The lattice secondary structures are local minima of the introduced energy.
\end{abstract}

\centerline{\sl Keywords: lattice polymers, secondary structures of proteins}

\section{Introduction}

In the present paper we construct a model of lattice protein which describes the formation of lattice secondary structures (alpha--helices and beta--sheets). We use lattice models of hydrogen bonds in the backbone of the peptide chain and lattice model of the Ramachandran plot which describes the geometric restrictions on the conformations of peptide chains. Before we describe our model let us mention some previous work.

For previous discussions of lattice models of polymers see, e.g. \cite{GrosbergKhokhlov,Grosberg,Dill2,ShakhGutin0,KlimovThirumalai}.
For a review of physics of proteins see \cite{FP} and for a discussion of the problem of protein folding see \cite{OnuchicWolynes}.
In \cite{Istrail} a review of applications of combinatorial algorithms to protein folding was given.

In \cite{BJST}, \cite{TBJ} lattice and off--lattice models of polymeric chains taking into account the directed hydrogen bond interaction were considered. The secondary structures were discussed in connection with the long range order in such models.

In the present paper we introduce a new lattice model of protein conformations. Our aim is to construct a lattice model which will approximate the geometry of conformations of real proteins, in particular, the geometry of secondary structures of proteins. We model a protein conformation by a continuous broken line in the lattice $\mathbb{Z}^3$. The $C_{\alpha}$--atoms of the peptide chain will lie at some vertices of the lattice $\mathbb{Z}^3$, backbone parts of amino acid residues (which connect the consecutive $C_{\alpha}$--atoms in the peptide chain) will correspond to the edges of the conformation of a lattice protein.

In the standard approach to lattice polymers the monomers in a polymer chains are described by edges of a lattice conformation which connect the nearest vertices in the lattice  $\mathbb{Z}^3$. The angles between such edges can be equal either $\pi/2$ or $\pi$. Therefore the standard approach to lattice polymers involves serious restrictions on the possible geometry of a polymer chain. In particular this geometry considerably differs from the geometry of peptide chains where the angles between the consecutive backbone residues usually vary from approximately $100^{\circ}$ (for  alpha--helices) to $120^{\circ}$ (for beta--sheets).

We propose a lattice model of protein conformations where backbone parts of amino acid residues are modeled by edges in the lattice $\mathbb{Z}^3$ which are not parallel to coordinate axes (in particular these edges will connect vertices of $\mathbb{Z}^3$ which are not neighbors).
The model is given by the set of possible edges, the set of possible angles between the edges and by the discrete analogue of the Ramachandran diagram (which describes the possible pairs of the backbone dihedral angles for $C_{\alpha}$--atoms in a peptide chain). In order to built the discrete Ramachandran plot we consider orientations of the planes of amino acid residues (as lattice vectors which are approximately orthogonal to the corresponding edges of the lattice model). The orientation of the edge describes the direction of the possible hydrogen bond.

Thus in the model under consideration a conformation of the backbone of the peptide chain is described by the sequence of edges of the lattice model, where each edge is characterized by the beginning, end and the orientation vectors. The rules of selection for the possible angles between the consecutive edges and the pairs of their orientations will realize a discrete model of the Ramachandran plot.

We show that the proposed model of lattice polymers allows to build lattice models of secondary structures of proteins such as alpha--helices and beta--sheets. Moreover these lattice secondary structures will have the correct structure of hydrogen bonds (described by the orientations vectors of the edges).

We also consider the side chains of amino acid residues (and the dependence of the Ramachandran plot on the side chain). We introduce an expression for the energy of a lattice protein which contains the contributions from the hydrogen bonds of the backbone and interaction of the side chains. Lattice secondary structures in the model under consideration can be considered as the minima of the energy of lattice polymers where the majority of hydrogen bonds in the backbone of the polymer chain are saturated and the geometric restrictions for conformations of the lattice peptide chains are satisfied.

The structure of the present paper is as follows.

In section 2 we describe the space of possible conformations for the model of a lattice polymer under consideration.

In section 3 we describe lattice secondary structures in the considered model, namely alpha--helices and beta--sheets.

In section 4 we construct the energy for our model of lattice polymer. The expression for the energy contains contributions from hydrogen bonds in the backbone of the chain and from interaction of side chains.

In section 5 we present some conclusion based on the results of the present paper.

\section{Conformations of lattice proteins}

In the present section we describe the set of conformations for the model of lattice proteins under consideration.

The backbone parts of amino acid residues will be modeled by some edges in the lattice $\mathbb{Z}^3$ which are not necessarily parallel to the coordinate axes. Here any edge of the lattice connects two vertices of the lattice. The cubic lattice $\mathbb{Z}^3$ is the group generated by the unit vectors $e_1$, $e_2$, $e_3$ (parallel to the coordinate axes), the edges correspond to linear combinations of these unit vectors with integer coefficients.

\medskip

The conformation of the backbone of a peptide chain for the model of lattice protein under consideration will correspond to the map $\Gamma:\{1,\dots,N\}\to\mathbb{Z}^3$ satisfying the conditions:

1) Any pair $i,i+1$ belonging to $\{1,\dots,N\}$ maps to vertices  $\Gamma(i)$, $\Gamma(i+1)$ from $\mathbb{Z}^3$ connected by an edge from the fixed set of edges, see below;

2) A contact between the consecutive edges $E_{i-1}=(\Gamma(i-1),\Gamma(i))$, $E_i=(\Gamma(i),\Gamma(i+1))$ of the chain should be allowed, see the definition \ref{Ramachandran} below;

3) A distance between any two edges of the lattice polymer is larger or equal to two.

\medskip

The vertices of the lattice polymer chain are ordered along the chain. This order fixes the signs of the coefficients of a vector corresponding to an edge --- a vector corresponds to the translation from the smaller to the larger vertex.

Let us consider the set of the following edges in the lattice $\mathbb{Z}^3$: edges connect vertices related by translations of the form
\begin{equation}\label{edge}
A_1e_1+A_2e_2+A_3e_3,\qquad A_i\in \{0,\pm 1, \pm 2\},
\end{equation}
where for each edge one of the coefficients $A_i$ is equal to zero, the second coefficient is equal to $\pm 1$, and the third one is equal to $\pm 2$ (the coefficients equal to $0,\pm 1, \pm 2$ can be selected in any order).

Analogously, the vectors (\ref{edge}) can be considered as arbitrary lattice rotations (combinations of rotations by $\pi/2$ with respect to the coordinate axes) of the vector
\begin{equation}\label{edge0}
2e_1+e_2.
\end{equation}

We will also consider some additional edges (which we call nonstandard) which are not of the form (\ref{edge}), namely corresponding to the lattice rotations of the vectors
\begin{equation}\label{edge2}
2e_1,\qquad 2e_1+e_2+e_3.
\end{equation}

In the following we will not distinguish the edges and the corresponding vectors.

An edge in the lattice gives only a partial description of the backbone part of the amino acid residue. One has to describe the orientation of an edge, i.e. the direction in which an edge can have hydrogen bonds.

\begin{definition}{\sl Let an edge $E$ correspond to some lattice rotation of a vector (\ref{edge0}) or (\ref{edge2}). The orientation of the edge $E$ is a vector $e$ equal to some vector from the basis $\{e_1,e_2,e_3\}$ of the lattice taken with the sign $\pm 1$, where $e$ is orthogonal to the vector with the largest (with the modulus two) coefficient in the expansion of $E$ over the basis $\{e_1,e_2,e_3\}$.}
\end{definition}

Therefore an edge with the orientation is described by the pair of vectors $(E,e)$, any edge $E$ possesses four possible orientations.

The cosine of the angle between the two edges of the lattice of the described above form can take values from some finite set. We will say that the angle between the two consecutive edges of the form (\ref{edge}) is allowed if the cosine of this angle takes values from the set $\{-0.6,-0.4,-0.2\}$ (where the angle between the consecutive edges $E$, $E'$ is the angle between vectors  $E$, $-E'$).

The allowed angles between the nonstandard edges (\ref{edge2}), and between the standard and nonstandard edges are described below in formulas (\ref{E0})--(\ref{E4}) è (\ref{E2'}).

\bigskip

\noindent{\bf Remark}\quad The angle between the covalent bonds at the $C_{\alpha}$--atom of peptide chain is equal approximately to $109^{\circ}$. In our model the angle between the consecutive edges is variable. This is related to the observation that the edge connects the two consecutive $C_{\alpha}$--atoms and is not parallel to the corresponding covalent bonds. The angle between the edges will vary with the rotation of the edges and depends on the corresponding dihedral angles. The angle between the edges belongs to some interval (approximately between $100^{\circ}$ and $120^{\circ}$, see for example \cite{FP}), in our discrete model the cosine of this angle will take values from the set  $\{-0.6,-0.4,-0.2\}$ (for standard edges (\ref{edge})).

\bigskip

We describe the side chain of the $i$-th amino acid residue (connected to the $i$-th $C_{\alpha}$--atom of the chain) with the help of a vector $s_i$ with the initial point in the $i$-th vertex of the lattice polymer. We denote by $(S_i,R_i)$ a pair (the position of the side chain in the lattice, the kind of the side chain). Thus the vector $s_i$ connects the vertex $\Gamma(i)$ of the lattice peptide chain with the point $S_i\in\mathbb{Z}^3$ where the side chain of the kind $R_i$ is situated.

\bigskip

Some examples of possible conformations $\Gamma$ of lattice proteins satisfying the described above conditions (without side chains) are shown at figures 1, 2, 3. Figures 1, 2 describe the examples of lattice alpha--helices and beta--sheets. Fig. 3 shows the left helix which can not be found in real proteins but satisfies the above conditions for edges and angles between the edges.

In order to eliminate the left helix and similar conformations one has to take into account the Ramachandran diagram (selection rules for possible pairs of dihedral angles at $C_{\alpha}$--atoms), see for example \cite{FP}. In our lattice model the discrete analogue of the Ramachandran diagram is defined by the selection rules for possible pairs of orientations for the consecutive edges of the chain.

The next definition describes the discrete Ramachandran plot and describes the positions of the corresponding side chains.

\begin{definition}\label{Ramachandran}
{\sl The list of allowed contacts for consecutive edges in the lattice polymer chain has the following form.

The consecutive edges $(E,e)$, $(E',e)$, where $E$, $E'$ have the form (\ref{edge})
\begin{equation}\label{EE'}
E=A_1e_1+A_2e_2+A_3e_3,\qquad E'=A'_1e_1+A'_2e_2+A'_3e_3
\end{equation}
with the orientations $e$, $e'$ have the allowed contact, if their form (\ref{EE'}) and orientations (up to lattice rotations) are described in one of the cases 1--6 below:

\medskip

\begin{itemize}
\item{ 1) The case of alpha--helix, see Fig. 1. The cosine of the angle between the edges is equal to $-0.2$, the orientations of the consecutive edges coincide:
$$
E=2e_2-e_3,\quad E'=2e_1-e_3,\quad e=e'=-e_3\quad s=-2e_1.
$$

}

\item{ 2) The case of beta--sheet, see Fig. 2. The cosine of the angle between the edges is equal to $-0.6$, the orientations of the consecutive edges are opposite:
$$
E=2e_1+e_2,\quad E'=2e_1-e_2,\quad e=e_3,\quad e'=-e_3,\quad s=2e_2.
$$
}
\end{itemize}

The cases 3--6 correspond to loops: the cosine of the angle between the edges is equal to $-0.4$, the orientations  $e$, $e'$ of the consecutive edges are orthogonal:
\begin{itemize}
\item{3) The cosine of the angle between the edges is equal to $-0.4$,
$$
E=2e_1+e_2,\quad E'=e_1 - 2 e_3,\quad e=e_2 ,\quad e'= -e_1\quad s=e_1+e_3.
$$

}

\item{4) The cosine of the angle between the edges is equal to $-0.4$,
$$
E=2e_1+e_2,\quad E'=e_1 + 2 e_3,\quad e=e_3 ,\quad e'= -e_2\quad s=2e_2.
$$
}
\end{itemize}

The next two cases should be allowed only for glycine.
\begin{itemize}
\item{5)  The cosine of the angle between the edges is equal to $-0.4$,
$$
E=2e_1+e_2,\quad E'=2e_2 - e_3,\quad e=e_2,\quad e'=e_1,\quad s=e_1-e_2.
$$
}

\item{6) The cosine of the angle between the edges is equal to $-0.4$,
$$
E=2e_1+e_2,\quad E'=2e_2 + e_3,\quad e=e_2,\quad e'=-e_3\quad s=e_1+e_3.
$$
}
\end{itemize}

The cases 7--10 of contacts of consecutive edges below describe contacts of standard and non--standard edges for beta turns (or beta bends).

The sequence of edges with orientations (up to lattice rotations) for a beta turn of the type 1 is described in formulas (\ref{E0})--(\ref{E4}), for a beta turn of the type 2 the formula (\ref{E2}) should be replaced by (\ref{E2'}) (where the edge $E_2$ has the opposite orientation).
The sequence of edges for a beta turn has the form $E_0E_1E_2E_3E_4$, where $E_0$ and $E_4$ belong to beta--strands, $E_1E_2E_3$ belong to the beta turn, see Fig. 4.
\begin{equation}\label{E0}
E_0=2e_1-e_2,\quad e^{(0)}=-e_3;
\end{equation}
\begin{equation}\label{E1}
E_1=2e_1+e_2+e_3,\quad e^{(1)}=e_3;
\end{equation}
\begin{equation}\label{E2}
E_2=2e_3,\quad e^{(2)}=e_2;
\end{equation}
\begin{equation}\label{E3}
E_3=-2e_1-e_2+e_3,\quad e^{(3)}=e_3;
\end{equation}
\begin{equation}\label{E4}
E_4=-2e_1+e_2,\quad e^{(4)}=-e_3.
\end{equation}
\begin{equation}\label{E2'}
E_2=2e_3,\quad e^{(2)}=-e_2.
\end{equation}
The edges $E_1E_2E_3$ which belong to the beta turn have a non--standard form which can not be described by formula (\ref{edge}).

The contacts for a beta turn have the form:

\begin{itemize}

\item{7) Contact of the edges $E_0$, $E_1$ given by (\ref{E0}), (\ref{E1}): $s_1=-2e_2$.}

\item{8) Contact of the edges $E_1$, $E_2$ given by (\ref{E1}), (\ref{E2}): $s_2=e_1-e_2-e_3$.}

\item{9) Contact of the edges $E_2$, $E_3$ given by (\ref{E2}), (\ref{E3}): $s_3=e_1-e_2+e_3$.}

\item{10) Contact of the edges $E_3$, $E_4$ given by (\ref{E3}), (\ref{E4}): $s_4=-2e_2$.}

\end{itemize}
}
\end{definition}

If the angles between the consecutive edges of the form (\ref{edge}) have the cosines $-0.2$ or $-0.6$, then the orientations of the edges are defined by the conformation of the polymer chain (as for real alpha--helices and beta--sheets). If the cosine of the angle between the consecutive edges of the form  (\ref{edge}) equals $-0.4$, then there are several possibilities for orientations of the edges. For glycine we have two additional possibilities for allowed contacts. Non--standard edges were used for the description of beta turns.

\begin{definition}\label{protein}
{\sl The set  $(\Gamma,\{e^{(i)}\})$ is a conformation of a lattice protein if:

i) $\Gamma$ is an embedding
$$
\Gamma:\{1,\dots,N\}\to\mathbb{Z}^3;
$$
where $i,i+1$ belonging to $\{1,\dots,N\}$ map to the vertices $\Gamma(i)$, $\Gamma(i+1)$ in $\mathbb{Z}^3$ connected by an edge $E_i$;

ii) the edges $E_i$, $i=1,\dots,N-1$ have the form of lattice rotations of (\ref{edge0}), (\ref{edge2})  and have some orientations $e^{(i)}$;

iii) the two consecutive edges for $\Gamma$ have the allowed contact in the sense of definition \ref{Ramachandran};

iv) the distance between any two edges of the lattice polymer and between the side chains is allowed (i.e. larger or equal to two).
}
\end{definition}

The introduced lattice model is based on the possibility to describe the geometry of the polymer chain (i.e. the angles between the edges corresponding to monomers) using the edges which connect vertices which are not neighbors in the lattice $\mathbb{Z}^3$, and on the discrete model of the Ramachandran plot. We will show that the defined model approximates well secondary structures of proteins. We will propose an expression for the energy of a conformation of lattice protein based on the description of hydrogen bonds between the backbone parts of the monomers and wills show that the minima of this energy will coincide with the lattice secondary structures.

\section{Secondary structures}

Let us describe lattice models of protein secondary structures and show that these structures are compatible with our model in the sense of definition \ref{protein}.  Figures 1 and 2 shows the lattice alpha--helix and beta--sheet (in projection). The alpha--helix can be directed along any of the coordinate axes, the beta--sheet can be parallel to any of the coordinate planes and beta--strands can be directed along any of the coordinate axes in this plane. Let us note that the (infinite) alpha--helix is invariant with respect to the following transformation --- the clockwise rotation by $90^{\circ}$ with the translation forward by one step of the lattice.

An edge of the lattice model describes a backbone part of a monomer (amino acid residue).  One turn of a lattice alpha--helix contains four edges ($3.6$ monomers in real proteins, \cite{FP}). The cosine of the angle between the consecutive edges in a lattice alpha--helix is equal to $-0.2$, and in a lattice beta--sheet it is equal to $-0.6$, which are close to the values (of the cosines of the angles) for real proteins. It is natural to take the step of the lattice $\mathbb{Z}^3$ (the distance between the nearest points in $\mathbb{Z}^3$)  to be equal to $1.5$ angstrom (this will give realistic sizes for the models of secondary structures, in particular for the distance between the turns in alpha--helix).

The lattice models shown at figures 1 and 2 are similar to the real secondary structures one finds in proteins. The orientations of the edges in lattice secondary structures have the following form. In alpha--helix the orientation vectors are parallel to the helix and are parallel to each other (this corresponds to the presence of hydrogen bonds between the edges obtained by a translation by a unit vector along the helix). In beta--sheet the orientation vectors are directed from the edges to the parallel edges in the parallel beta--strands, the orientation vectors are antiparallel for the consecutive edges in the chain (which again reflects the picture of hydrogen bonds between the strands in a beta--sheet).

The orientations of edges are parallel to the coordinate axes, thus in the model under consideration the secondary structures will be parallel to the coordinate axes, a feature which is natural for lattice models.

\begin{figure}
\begin{pspicture}(4,5)
   \psline[linewidth=0.5pt](0,0)(0.2,3.2)(3.4,3.4)(3.6,0.6)(0.8,0.8)(1,4)(4.2,4.2)(4.4,1.4)(1.6,1.6)(1.8,4.8)(5,5)(5.2,2.2)(2.4,2.4)
   \rput(6,0){$\alpha$--helix}
\end{pspicture}
\rput(3,2){Fig. 1}
\end{figure}

\begin{figure}
\begin{pspicture}(8,5)
   \psline[linewidth=0.5pt](0,1)(3,2.5)(6,1)(9,2.5)(12,1)
   \psline[linewidth=0.5pt](0.8,1.8)(3.8,3.3)(6.8,1.8)(9.8,3.3)(12.8,1.8)
   \psline[linewidth=0.5pt](1.6,2.6)(4.6,4.1)(7.6,2.6)(10.6,4.1)(13.6,2.6)
   \psline[linewidth=0.5pt](2.4,3.4)(5.4,4.9)(8.4,3.4)(11.4,4.9)(14.4,3.4)
\end{pspicture}
\rput(2,1){$\beta$--sheet}
\rput(2,0){Fig. 2}
\end{figure}

\begin{figure}
\begin{pspicture}(4,5)
   \psline[linewidth=0.5pt](0,0)(3.2,0.2)(3.4,3.4)(0.6,3.6)(0.8,0.8)(4,1)(4.2,4.2)(1.4,4.4)(1.6,1.6)(4.8,1.8)(5,5)(2.2,5.2)(2.4,2.4)
   \rput(6,0){Left helix}
\end{pspicture}
\rput(3,2){Fig. 3}
\end{figure}

\begin{figure}
\begin{pspicture}(8,3)
   \psline[linewidth=0.5pt](3,2.5)(6,1)(9.2,2.7)(9.6,3.1)(6.8,1.8)(3.8,3.3)
\end{pspicture}
\rput(2,1){$\beta$--turn}
\rput(2,0){Fig. 4}
\end{figure}

\begin{figure}
\begin{pspicture}(4,8)
   \psline[linewidth=0.5pt](0,3)(0.2,6.2)(-2.8,6.2)(0.2,6.2)(3.4,6.4)(3.4,9.4)(3.4,6.4)(3.6,3.6)(6.6,3.6)(3.6,3.6)(0.8,3.8)(0.8,0.8)(0.8,3.8)
   (1,7)(-2,7)(1,7)(4.2,7.2)(4.2,10.2)(4.2,7.2)(4.4,4.4)(7.4,4.4)(4.4,4.4)(1.6,4.6)(1.6,1.6)(1.6,4.6)(1.8,7.8)(-1.2,7.8)(1.8,7.8)
   (5,8)(5,11)(5,8)(5.2,5.2)(8.2,5.2)(5.2,5.2)(2.4,5.4)
   \rput(10,1){$\alpha$--helix with side chains}
\end{pspicture}
\rput(3,2){Fig. 5}
\end{figure}

\begin{figure}
\begin{pspicture}(8,8)
   \psline[linewidth=0.5pt](0,4)(3,5.5)(3,8.5)(3,5.5)(6,4)(6,1)(6,4)(9,5.5)(9,8.5)(9,5.5)(12,4)
   \psline[linewidth=0.5pt](0.8,4.8)(3.8,6.3)(3.8,9.3)(3.8,6.3)(6.8,4.8)(6.8,1.8)(6.8,4.8)(9.8,6.3)(9.8,9.3)(9.8,6.3)(12.8,4.8)
   \psline[linewidth=0.5pt](1.6,5.6)(4.6,7.1)(4.6,10.1)(4.6,7.1)(7.6,5.6)(7.6,2.6)(7.6,5.6)(10.6,7.1)(10.6,10.1)(10.6,7.1)(13.6,5.6)
   \psline[linewidth=0.5pt](2.4,6.4)(5.4,7.9)(5.4,10.9)(5.4,7.9)(8.4,6.4)(8.4,3.4)(8.4,6.4)(11.4,7.9)(11.4,10.9)(11.4,7.9)(14.4,6.4)
\end{pspicture}
\rput(3,2){$\beta$--sheet with side chains}
\rput(0,1){Fig. 6}
\end{figure}

Figures 5 and 6 show the lattice alpha--helix and beta--sheet with side chains.

\section{A model of energy of a lattice protein}

Secondary structures in proteins are stabilized by hydrogen bonds in the backbone of the peptide chain. Any amino acid residue can obtain two hydrogen bonds which should be approximately orthogonal to the backbone of the peptide chain.

Let us consider the translations of an edge $E$ by $\pm 4 e$ (where $e$ is the unit vector of orientation of $E$). We will say that an edge $E$ with orientation $e$ can have hydrogen bonds with the described translation if the orientations of the edge $E$ and its translation coincide.

We introduce the expression for the energy of a conformation of a lattice protein which contains the contributions from hydrogen bonds in the backbone of the chain and from the interaction between the side chains of amino acid residues.

A conformation of a lattice protein of the length $N$ is defined by the map $$\Gamma:\{1,\dots,N\}\to\mathbb{Z}^3$$ satisfying the set of conditions of definition \ref{protein}. The edge $E_i=(\Gamma(i),\Gamma(i+1))$ corresponds to the translation from $\Gamma(i)$ to $\Gamma(i+1)$, the orientation of this edge will be denoted by  $e^{(i)}$.

We introduce the energy of the conformation $\Gamma$ of a lattice protein as follows
\begin{equation}\label{E_HB}
E(\Gamma)=E_{HB}(\Gamma)+E_{SC}(\Gamma)=$$ $$
=-\sum_{i,j=1}^{N-1}\delta(E_i+4e^{(i)},E_j)\delta(e^{(i)},e^{(j)})+\sum_{i=1}^{N-1}p(E_i)-\sum_{i,j=2}^{N-1}\theta(S_i,S_j)M_{R_i,R_j}.
\end{equation}

The first sum in the expression above describes the contribution to the energy from hydrogen bonds in the backbone of the peptide chain. Here $\delta(E_i+4e^{(i)},E_j)$ is equal to one when the edges $E_j$ and $E_i+4e^{(i)}$ (the translation of $E_i$ along the orientation vector $e^{(i)}$) coincide and is equal to zero otherwise. Note that the edges $E_j$ and $E_i+4e^{(i)}$ should coincide as sets (i.e. the corresponding vectors might be antiparallel). Analogously $\delta(e^{(i)},e^{(j)})$ is equal to one when the orientation vectors of the $i$-th and the $j$-th edges coincide (i.e. are parallel) and is equal to zero otherwise.

The second sum in (\ref{E_HB}) describes energetic penalties for the use of non--standard edges. The energetic penalty $p(E_i)$  is equal to zero if the edge $E_i$ is standard and is equal to some positive value otherwise. Nonstandard edges are involved in beta--turns where energetic penalties are related to the torsion of the peptide chain. Energetic penalties (say for beta--turns in beta--sheets) can be compensated by the energies of hydrogen bonds and interaction of side chains.

For the edges $E_1$ and $E_3$ of the beta--turn, see (\ref{E0})--(\ref{E4}), there exists a hydrogen bond. We ignore this bond in our model (this can be considered as a variant of energetic penalty).

The third sum in (\ref{E_HB}) describes the contribution to the energy of a lattice protein from the interaction of side chains of amino acid residues. Here $S_i$ and $R_i$ are the position and the type of the $i$-th side chain, the function $\theta(S_i,S_j)$ depends on the distance between the side chains $S_i$ and $S_j$ (for example is equal to one for distances not larger than 4 in the lattice and decreases to zero for a distance larger than 5), the matrix $M_{R_i,R_j}$ (the Miyazawa--Jernigan matrix \cite{MJ}) describes the interaction of side chains (the indices of this matrix  enumerate the set of 20 amino acids).

\bigskip

\noindent
{\bf Remark}\quad Let us discuss the lattice aplha--helix, see Fig. 5, and beta--sheet, see Fig. 6. It is easy to see that for these lattice conformations all contributions to the energy (\ref{E_HB}) related to hydrogen bonds in the backbone of the lattice peptide chain (excluding the beta--turns) are present in the first sum in (\ref{E_HB}), i.e. any edge excluding the beta--turns possesses two hydrogen bonds. This corresponds to the known property that all hydrogen bonds for the backbone of the peptide chain in secondary structures are saturated. Here we have to take into account the boundary effects -- for edges at the boundaries of the alpha--helix and beta--sheet only one hydrogen bond per edge is saturated.

Moreover the pairs of the nearest side chains of amino acid residues for lattice secondary structures have the distance four between the side chains, therefore these side chains interact (i.e. give contributions to the energy (\ref{E_HB})).

Therefore the energy (\ref{E_HB}) of lattice secondary structures (alpha--helices and beta--sheets) is low. For short lattice polymer chains the considered lattice secondary structures will be the minima of energy, for long lattice polymer chains combinations of interacting secondary structures (say of two parallel helices) will have the lower energy due to the interaction between the side chains of the different secondary structures.

It would be interesting to investigate the following question: do there exist alternative lattice secondary structures, i.e. conformations of lattice polymers which consist mainly of standard edges (\ref{edge}), such that for any standard edge in the conformation both hydrogen bonds are saturated (again, taking into account the boundary effects).

\section{Conclusion}

In the present paper we have constructed a lattice model for the conformation of a protein. For this model the main secondary structures of proteins (alpha--helices and beta--sheets) are local minima of the energy of a lattice polymer. We approximate the geometry of peptide chains using edges in the lattice $\mathbb{Z}^3$ which are not parallel to the coordinate axes and take into account the discrete form of the Ramachandran plot.

The constructed model is able to describe the hydrogen bonds in the backbone of a peptide chain. In particular the lattice alpha--helices and beta--sheets have the correct picture of hydrogen bonds. The model also takes into account the interaction of side chains.

In this model the local minima of the energy (\ref{E_HB}) take the form of combinations of secondary structures since this is the only way to saturate a large number of hydrogen bonds in the backbone of the peptide chain and satisfy the geometric restrictions on the form of the peptide chain (as for real proteins).

\bigskip

\noindent{\bf Acknowledgments}\qquad This work is partially supported by the DFG project AL 214/40-1.
One of the authors (S.K.) gratefully
acknowledges being partially supported by the grants of
the Russian Foundation for Basic Research
RFBR 11-01-00828-a and 11-01-12114-ofi-m-2011, by the grant of the President of Russian
Federation for the support of scientific schools NSh-2928.2012.1,  and
by the Program of the Department of Mathematics of the Russian
Academy of Science ``Modern problems of theoretical mathematics''.

\end{document}